\documentclass[preprintnumbers]{revtex4}
\usepackage{amsfonts}
\usepackage{amsmath}
\usepackage{amssymb}
\usepackage{graphicx}
\usepackage[cp1251]{inputenc}
\usepackage[russian]{babel}

\input{tcilatex}

\begin{document}

\title[Lifshitz-type thin films ]{Order parameter configurations in the
Lifshitz-type incommensurate ferroelectric thin films}
\author{Sergey A. Ktitorov}
\affiliation{A.F. Ioffe Physicotechnical Institute of the Russian
Academy of Sciences, Polytechnicheskaja str. 26, St. Petersburg
194021,\ Russia; St. Petersburg Electrotechnical University, prof.
Popov str. 5, St. Petersburg 197376, Russia}
\author{Philip A. Pogorelov}
\affiliation{St. Petersburg State University, Pervogo Maya str. 100, Petrodvoretz, St.
Petersburg 198504, Russia}

\begin{abstract}
The Dzialoshinskii model of periodic and helicoidal structures has
been analyzed without neglecting of the amplitude function
oscillations. The amplitude function oscillations are shown to be
important for understanding of the nature of the phase function.
Analytic consideration is carried out in the limit of small
anisotropy (neglecting the cosine term in the Hamiltonian).
Surprisingly, the phase jumps survive even in the limit of the
vanishing anisotropy.
\end{abstract}

\maketitle

\section{Introduction}

The theory of modulated structures in macroscopic ferromagnets \cite%
{dzialoshin}, ferroelectrics \cite{levaniuk} and metal alloys
\cite{hachatur} has a rich history. The number of works dedicated
to a study of the modulated structures in small samples is
significantly less \cite{char}. The standard approach to the
theory of the modulated structures (including the incommensurate
ones) uses the constant order parameter amplitude function
approximation \cite{levaniuk}, \cite{izium}. Our consideration of
the bounded sample (thin film, in particular) requires us to drop
this approximation: the order parameter must satisfy some boundary
conditions and it is unlikely to find a solution with constant
order parameter satisfying these conditions. That is why we have
to consider a generic case of connected set of the nonlinear
equations for the order parameter amplitude and phase functions.
We will show here that the solutions of this system are
significantly more complicated, than in the constant order
parameter approximation.

\section{Free energy and the equations for the order parameter for a
ferroelectric with the Lifshitz invariants}

\qquad The Landau free energy functional reads \cite{izium}:%
\begin{eqnarray}
\Phi & =\int dz\left\{
-r(\eta_{1}^{2}+\eta_{2}^{2})+u_{1}(\eta_{1}^{2}+\eta_{2}^{2})^{2}+u_{2}(%
\eta_{1}^{2}\eta_{2}^{2})\right\} +  \notag \\
& +\int dz\left\{ \sigma\left( \eta_{2}\frac{\partial\eta_{1}}{\partial z}%
-\eta_{1}\frac{\partial\eta_{2}}{\partial z}\right) +\gamma\left[ \left(
\frac{\partial\eta_{1}}{\partial z}\right) ^{2}+\left( \frac{\partial
\eta_{2}}{\partial z}\right) ^{2}\right] \right\} ,  \label{free}
\end{eqnarray}
where $\eta_{1}$ and $\eta_{2}$\ are the order parameter components; we
consider only one-dimensional configurations; parameters $r,$\ $\sigma$, $%
u_{1},u_{2}$ and $\gamma$ are the Landau free energy expansion coefficient.
Introducing the amplitude and phase variables

\begin{equation*}
\eta_{1}=\rho\cos\varphi,\eta_{2}=\rho\sin\varphi,
\end{equation*}
we obtain the following expression for the Landau free energy:%
\begin{eqnarray}
\Phi=\int dz \{ -r\rho^{2}+u\rho^{4}+w\rho^{n}(1+\cos n\varphi)  \notag \\
-\sigma\rho^{2}\frac{\partial\varphi}{\partial z}+ \gamma\left[ \left( \frac{%
\partial\rho}{\partial z}\right) ^{2}+\rho^{2}\left( \frac {\partial\varphi}{%
\partial z}\right) ^{2}\right] \} .  \label{free2}
\end{eqnarray}
Varying the free energy we obtain the equilibrium equations \cite{izium}%
\begin{eqnarray}
-r\rho+2u\rho^{3}+\frac{n}{2}w\rho^{n-1}(1+\cos n\varphi)+\gamma\rho (\frac{%
\partial\varphi}{\partial z})^{2}-  \notag \\
\gamma\frac{\partial^{2}\rho }{\partial z^{2}}-\sigma\rho\frac{%
\partial\varphi}{\partial z}=0,   \label{ampeq}
\end{eqnarray}

\begin{eqnarray}
\gamma\rho^{2}\frac{\partial^{2}\varphi}{\partial z^{2}}+2\gamma\rho \frac{%
\partial\rho}{\partial z}\frac{\partial\varphi}{\partial z}-\sigma \rho\frac{%
\partial\rho}{\partial z}+\frac{n}{2}w\rho^{n}\sin n\varphi=0.
\label{phaseeq}
\end{eqnarray}

Here $n$ is an integer number describing the system symmetry. Now we
introduce the dimensionless variables:%
\begin{eqnarray}
\rho=\sqrt{\frac{r}{2u}}R,\xi=z\sqrt{\frac{r}{\gamma}},\frac{dR}{dz}=\frac {%
dR}{d\xi}\sqrt{\frac{r}{\gamma}},  \notag \\
\frac{\sigma}{\sqrt{\gamma r}}=T,u^{1-\frac{n}{2}}nwr^{\frac{n}{2}-2}2^{-%
\frac{n}{2}}=K.   \label{dimlessvariables}
\end{eqnarray}
Then the equations (\ref{ampeq}) and (\ref{phaseeq}) take the form%
\begin{eqnarray}
R^{\prime\prime}-R^{3}+(1-\varphi^{\prime2}+T\varphi^{\prime})R-  \notag \\
R^{n-1}K(\cos n\varphi+1)=0,  \label{ampeqdimless}
\end{eqnarray}
\begin{eqnarray}
\varphi^{\prime\prime}+2\frac{R^{\prime}}{R}\varphi^{\prime}-\frac{R^{\prime}%
}{R}T+R^{n-2}K\sin n\varphi=0.   \label{phaseeqdimless}
\end{eqnarray}

\section{Approximate analytic solution}

Let us begin from the limit $K=0.$ Then the equation (\ref{phaseeqdimless})
can be solved:%
\begin{eqnarray}
\varphi^{\prime}\equiv\psi=\frac{C_{0}}{R^{2}}+\frac{T}{2},
\label{phasesolution}
\end{eqnarray}
where $C_{0}=\left[ \psi\left( 0\right) -\frac{T}{2}\right] R\left( 0\right)
^{2}$ is the integration constant, which is determined by the initial
conditions $\psi\left( 0\right) $ and $R\left( 0\right) $. Now we can
substitute this expression for $\varphi^{\prime}$ into the equation (\ref%
{ampeqdimless}). We obtain in result a closed equation for the amplitude
function $R:$%
\begin{eqnarray}
R^{\prime\prime}-R^{3}+R(1+\frac{T^{2}}{4})-\frac{C_{0}^{2}}{R^{3}}=0.
\label{ampclosed}
\end{eqnarray}

This equation can be interpreted as a dynamics equation with the effective
potential:%
\begin{eqnarray}
U=\frac{R^{2}}{2}(1+\frac{T^{2}}{4})-\frac{R^{4}}{4}+\frac{C_{0}^{2}}{2R^{2}}%
.   \label{effpot}
\end{eqnarray}

An interesting feature of this potential is its dependence on the initial
conditions via the constant $C_{0}^{2}.$ There exists a domain of
parameters, where the potential has a minimum and, therefore, an oscillating
solution for the amplitude function can take place. A condition of the
maximum and minimum points merging into the inflection point with the
horizontal derivative $U^{\prime}\left( R\right) =U^{\prime\prime}\left(
R\right) =0$ gives the equation

\begin{equation*}
T^{6}+12T^{4}+48T^{2}+64-432C_{0}^{2}=0
\end{equation*}

However, a presence of a minimum is necessary but not sufficient condition:
an oscillating solution will not exist if the initial point is situated
outside the potential well.

A border line for the domain, where oscillating solutions can exist is
depicted here:

\begin{center}
\includegraphics[
height=2in, width=1.5in, angle=270, trim = 0.1in 0.1in 0.1in 0.1in, clip
]{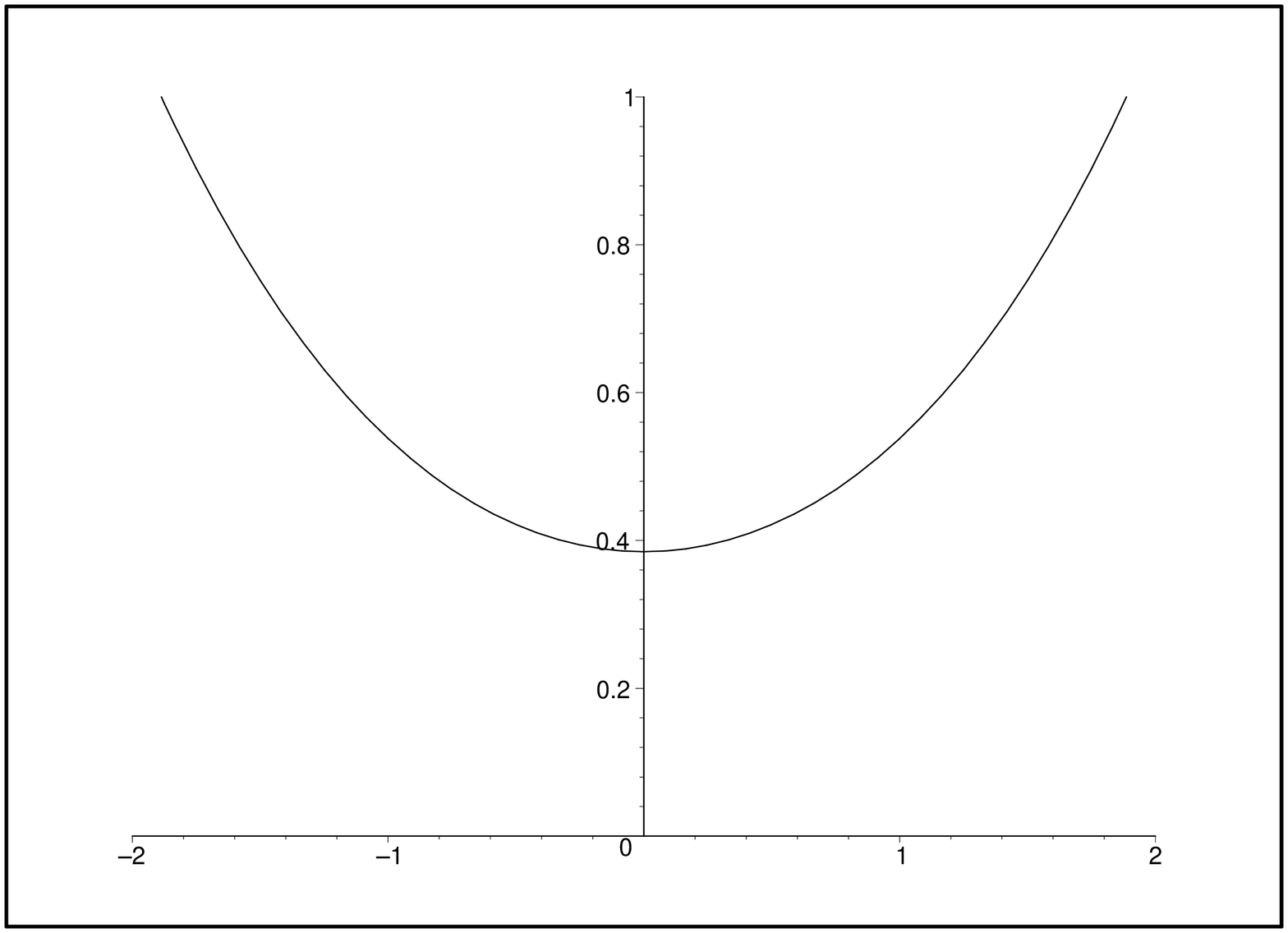}

FIG.1. A border line for the domain, where oscillating solutions can exist.
\end{center}

\bigskip

If we have a solution for the amplitude function $R(\xi),$ the phase
function $\varphi(\xi)$ is given by the equation (\ref{phasesolution}). The
phase function consists from two terms: a slow contribution stemming from
the second term in (\ref{phasesolution}) and more or less diffused jumps due
to the first term in this equation. It is important that the jump value
equals exactly to $\pi:$

The first integral of \ref{ampeqdimless} reads $\frac{dR}{d\xi}=\sqrt {2(E-U)%
},$ so $d\xi=\frac{dR}{\sqrt{2(E-U)}}.$

Notice that the jump value $\int\frac{C_{0}}{R^{2}}d\xi$ calculated in the
vicinity of the $R(\xi)$ minimum equals to
\begin{eqnarray}
\Delta\varphi & =\int\frac{C_{0}}{R^{2}}d\xi=\int\frac{C_{0}}{R^{2}}\frac {dR%
}{\sqrt{2(E-U)}}\simeq2\int_{R_{0}}^{R_{e}}\frac{C_{0}dR}{R^{2}\sqrt{2(\frac{%
C_{0}^{2}}{2R_{0}^{2}}-\frac{C_{0}^{2}}{2R^{2}})}}=  \notag \\
& =2\int_{R_{0}}^{R_{e}}\frac{dR}{R\sqrt{\frac{R^{2}}{R_{0}^{2}}-1}}%
=2\arccos\left\vert \frac{R_{0}}{Re}\right\vert \simeq\pi,
\label{phasejump}
\end{eqnarray}
where $R_{e}\gg R_{0}.$ We have used the first integral of
(\ref{ampclosed}) above. Only leading terms of $U(R)$ were taken
into account. Note that the model (\ref{free2}), which can be
reduced to the sine-Gordon model in the limit of $R=const$, admits
solutions with jumps exactly equal to $\pm\pi$ (topological charge
\cite{soliton}). However, we have obtained a similar
result within the approximation $K=0$, i.e. neglecting the cosine term in (%
\ref{free2})! As it is seen from the formulae \ref{phasejump}, these phase
jumps appear due to the excursion of the amplitude function near the
singularity $\frac{C_{0}}{R^{2}}$.

The equation (\ref{ampclosed}) can be solved analytically, but we
will present below numerical solutions both for $K=0$ and
$K\neq0.$

\section{Numerical solution for K = 0}

Some results of calculation of the amplitude and phase functions spatial
dependence for the case of $K=0$ are presented below. Figures 2 -- 5 and 6
-- 9 differ only by initial values of the amplitude function: in the case of
figures 2 -- 5 we have small initial amplitude function and, therefore,
oscillations of the amplitude function are near to harmonic ones, while in
the case of figures 6 -- 9 the initial amplitude function is near to the
apex of the hump and, therefore, we have a train of solitons. \newpage

\begin{center}
\includegraphics[
height=2in, width=1.5in, angle=270, trim = 0.1in 0.1in 0.1in 0.1in, clip
]{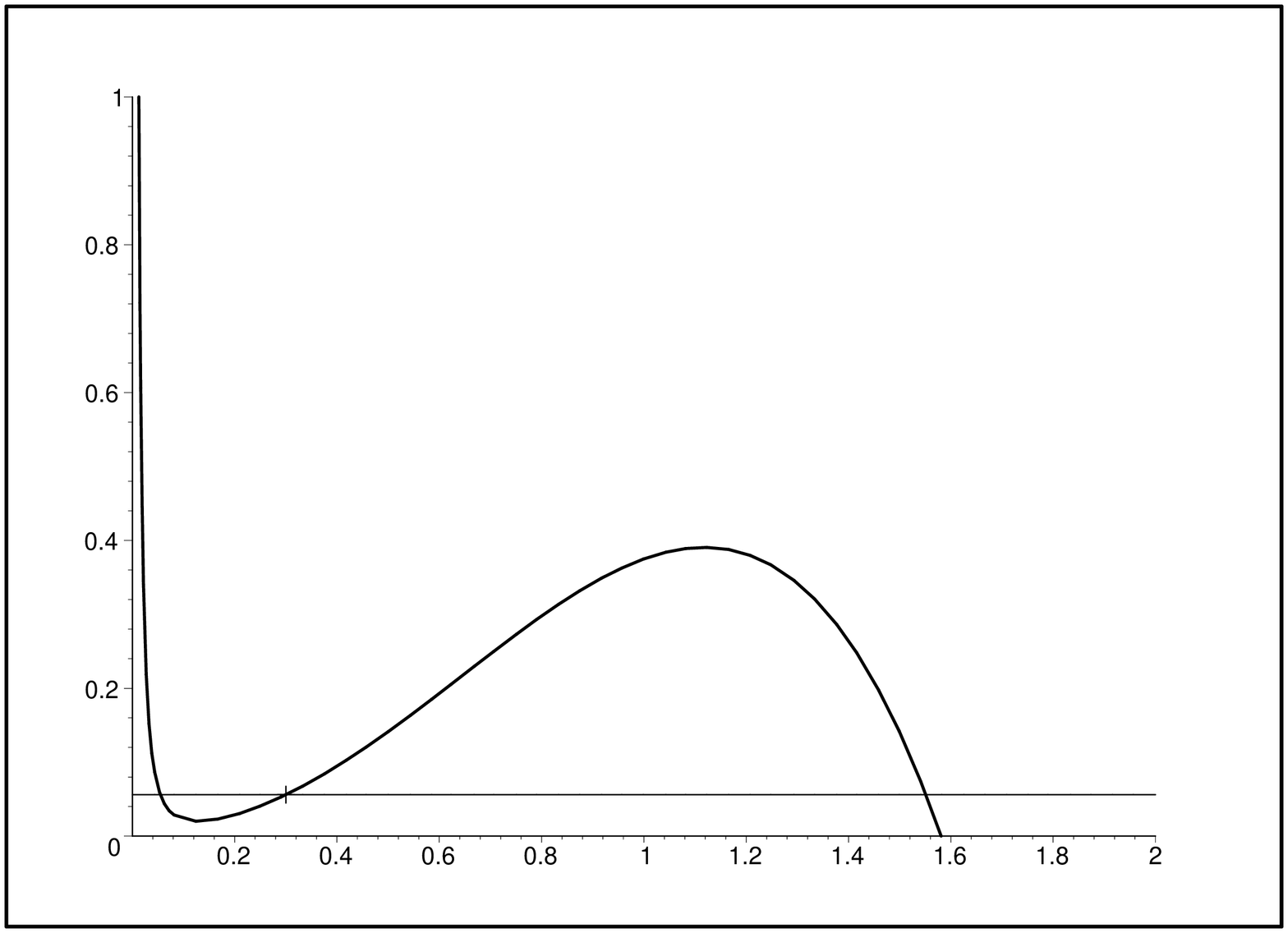}

FIG.2. Effective potential for $n=4,$ $K=0,$ $T=1,$ $R(0)=0.3,$ $R^{\prime
}(0)=0,$ $\varphi(0)=0,$ $\varphi^{\prime}(0)=0.3$.

Vertical dash marks $R(0)$ value.

\includegraphics[
height=2in, width=1.5in, angle=270, trim = 0.1in 0.1in 0.1in 0.1in, clip
]{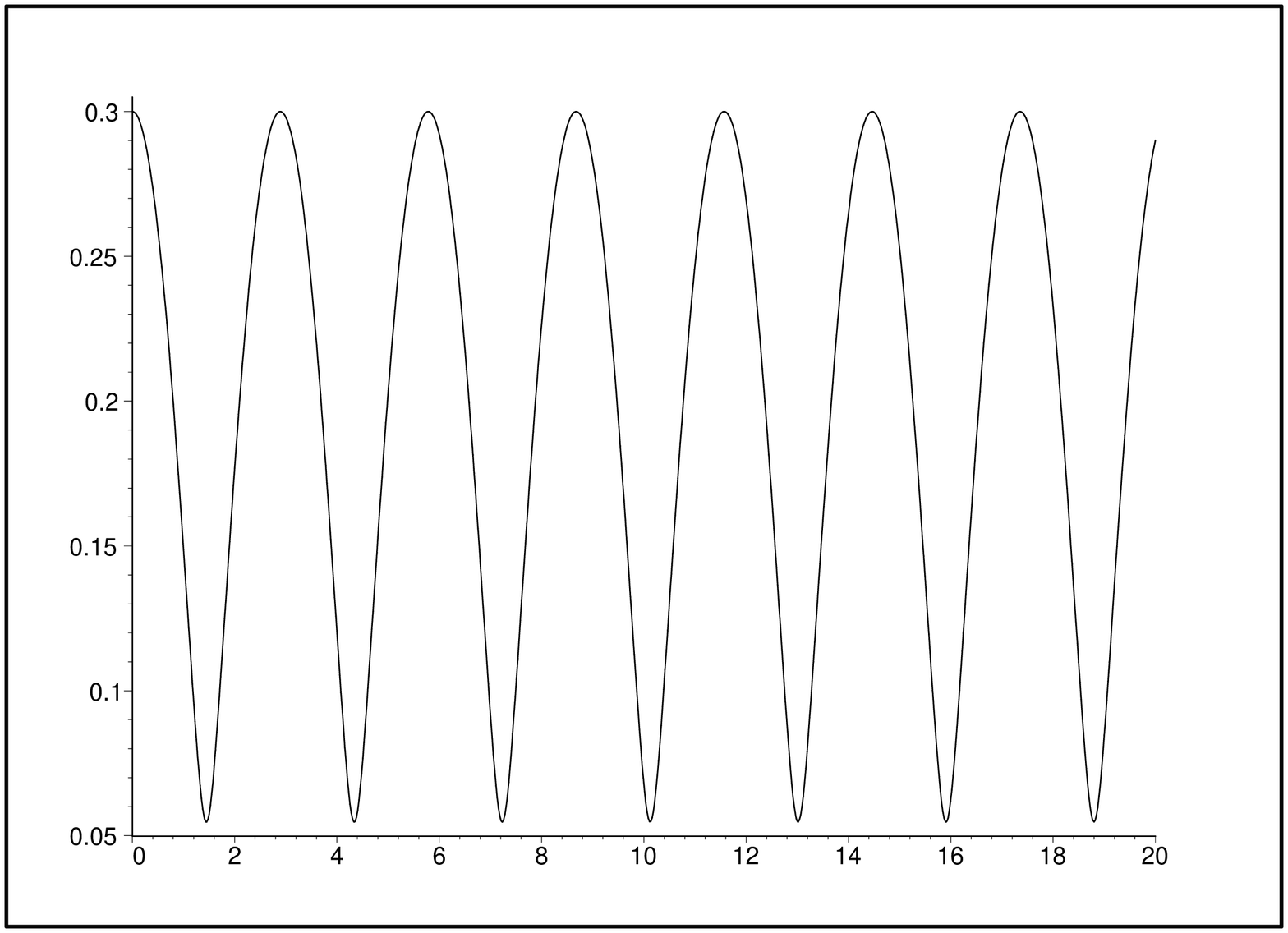}

FIG. 3. Spatial dependence of the amplitude function for $n=4,$ $K=0,$ $T=1,$
$R(0)=0.3,$ $R^{\prime}(0)=0,$ $\varphi(0)=0,$ $\varphi^{\prime}(0)=0.3$

\includegraphics[
height=2in, width=1.5in, angle=270, trim = 0.1in 0.1in 0.1in 0.1in, clip
]{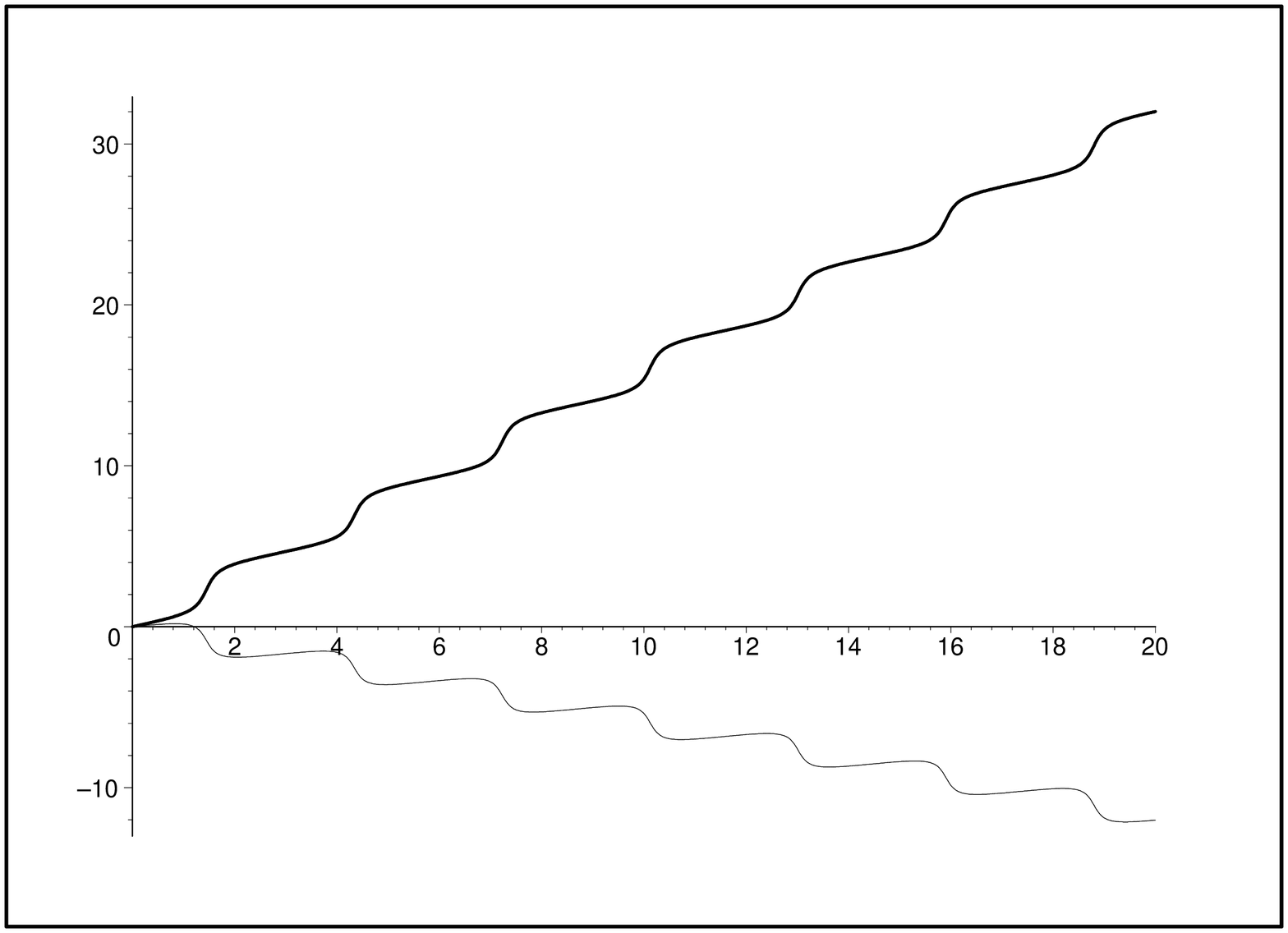}

FIG. 4. Spatial dependence of the phase function for $n=4,$ $K=0,$ $T=1,$ $%
R(0)=0.3,$ $R^{\prime}(0)=0,$ $\varphi(0)=0$

(a) thin line $\varphi^{\prime}(0)=0.3$

(b) heavy line $\varphi^{\prime}(0)=0.7.$

\includegraphics[
height=1.8in, width=1.8in, angle=270, trim = 0.1in 0.1in 0.1in 0.1in, clip
]{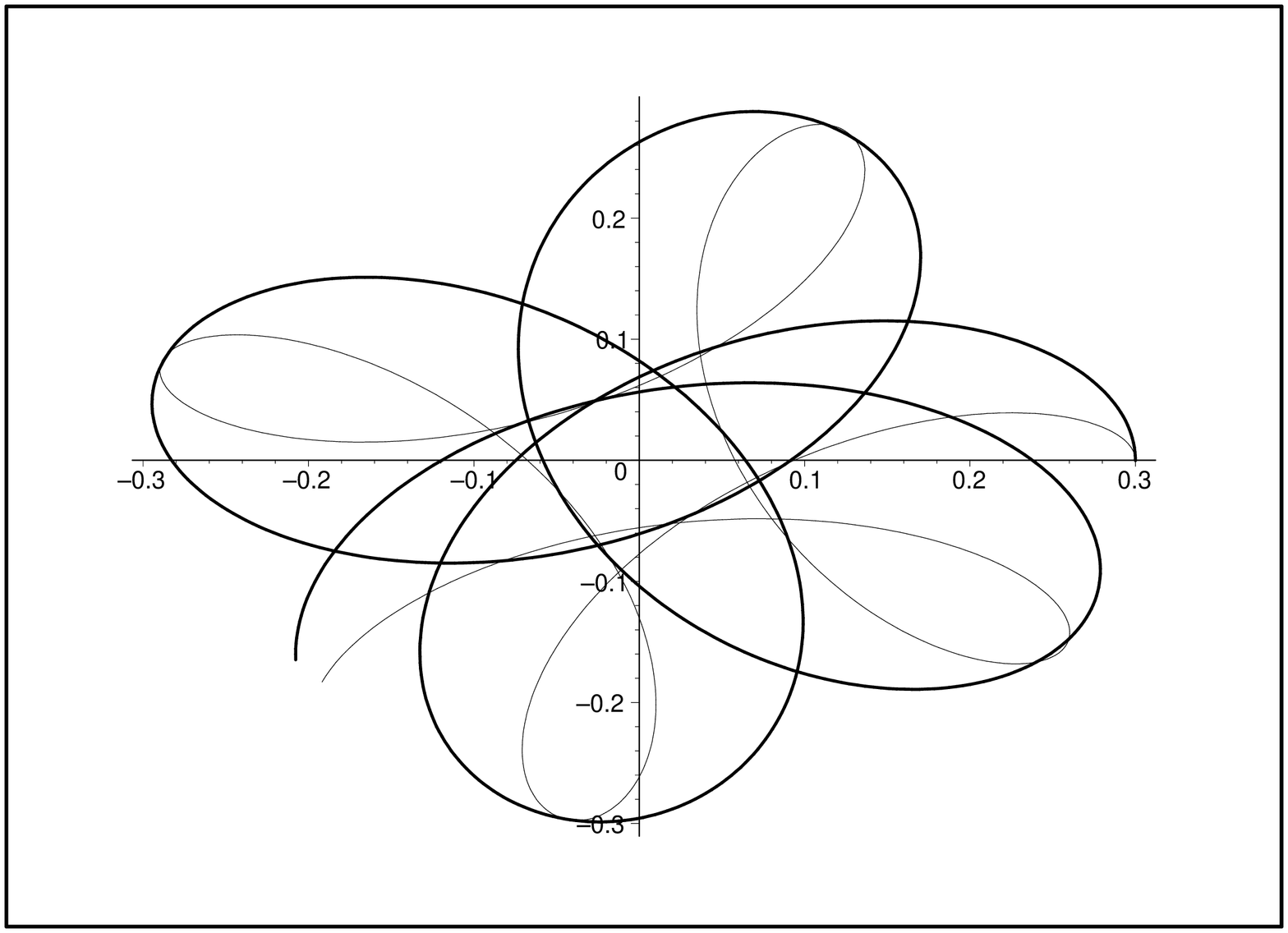}

FIG. 5. Amplitude-phase polar diagram for $n=4,$ $K=0,$ $T=1,$ $R(0)=0.3,$ $%
R^{\prime}(0)=0,$ $\varphi(0)=0$

(a) thin line $\varphi^{\prime}(0)=0.3$

(b) heavy line $\varphi^{\prime}(0)=0.7.$
\end{center}

\newpage

\begin{center}
\includegraphics[
height=2in, width=1.5in, angle=270, trim = 0.1in 0.1in 0.1in 0.1in, clip
]{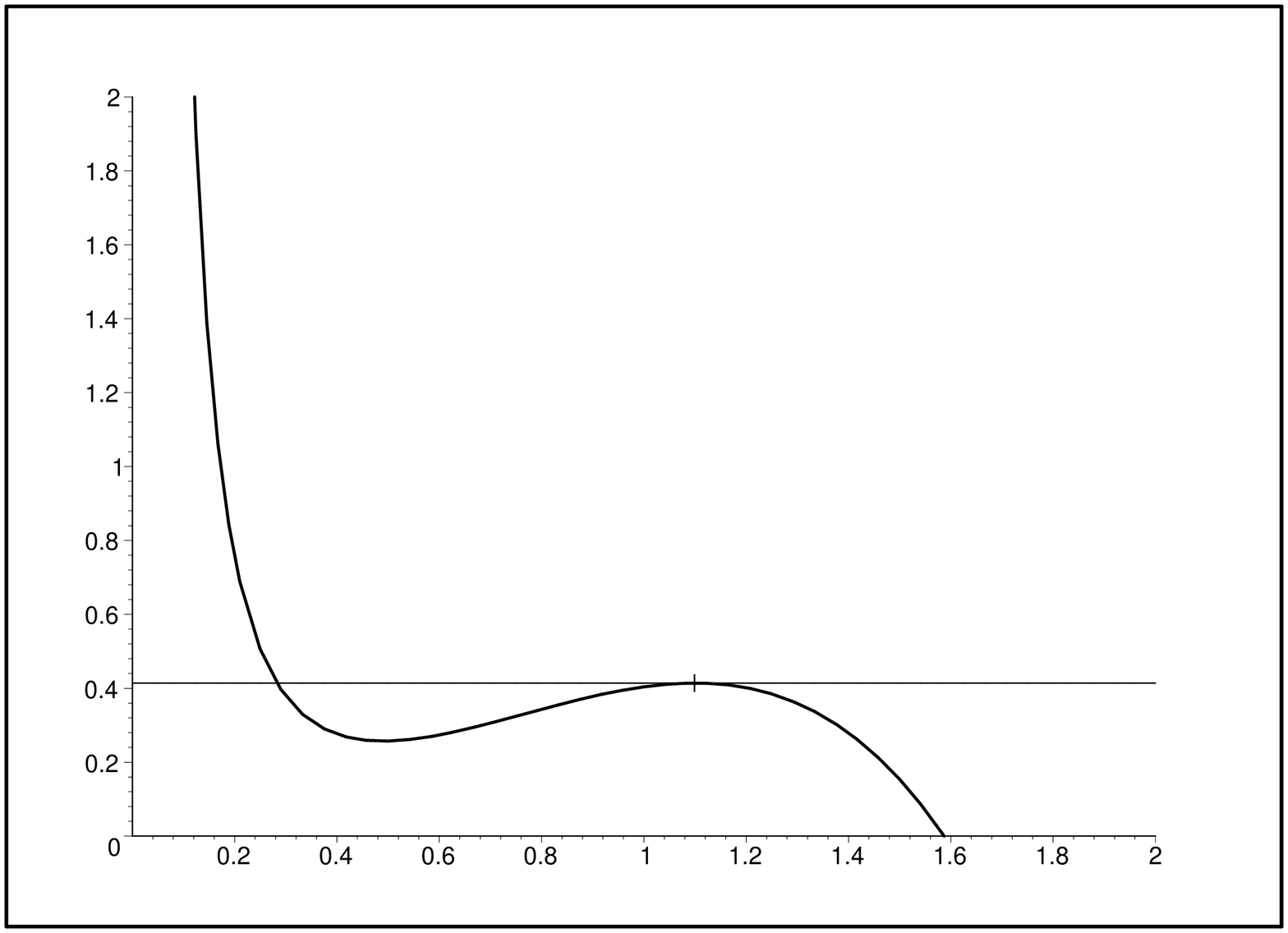}

FIG. 6. Effective potential for $n=4,$ $K=0,$ $T=1,$ $R(0)=1.099,$ $%
R^{\prime }(0)=0,$ $\varphi(0)=0,$ $\varphi^{\prime}(0)=0.3$

Vertical dash marks $R(0)$ value.

\includegraphics[
height=2in, width=1.5in, angle=270, trim = 0.1in 0.1in 0.1in 0.1in, clip
]{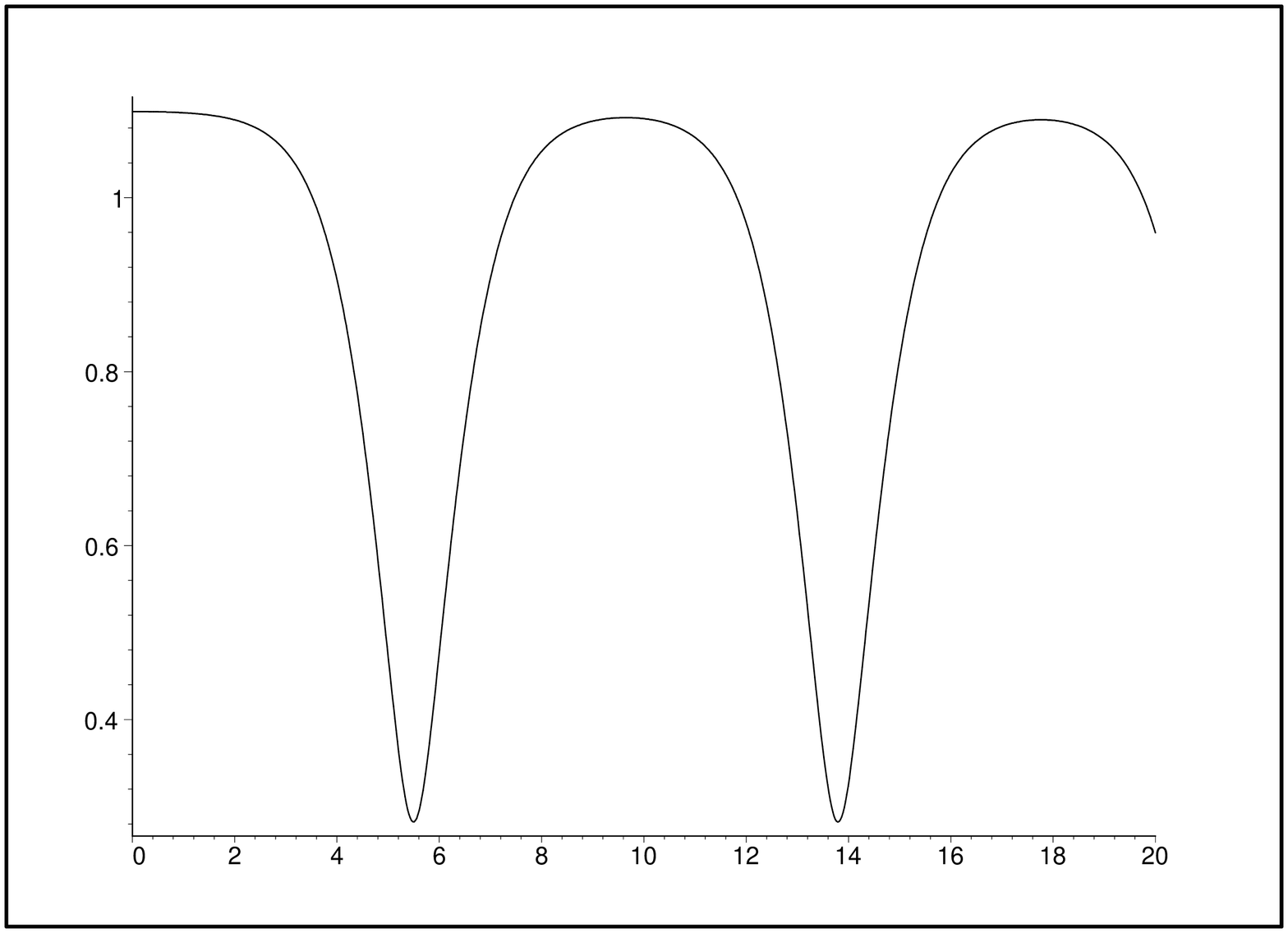}

FIG. 7. Spatial dependence of the amplitude function for $n=4,$ $K=0,$ $T=1,$
$R(0)=1.099,$ $R^{\prime}(0)=0,$ $\varphi(0)=0,$ $\varphi^{\prime}(0)=0.3$

\includegraphics[
height=2in, width=1.5in, angle=270, trim = 0.1in 0.1in 0.1in 0.1in, clip
]{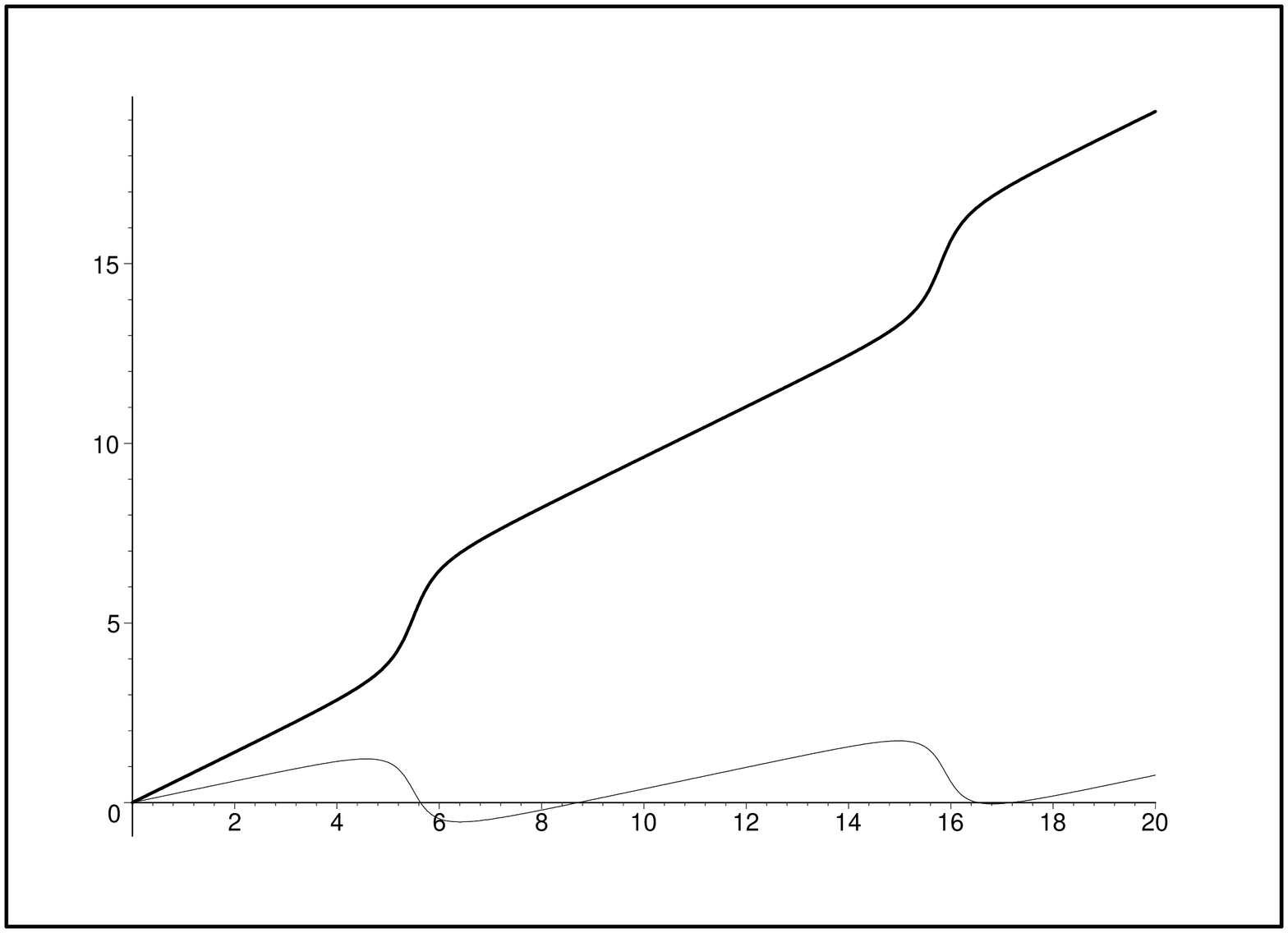}

FIG. 8. Spatial dependence of the phase function for $n=4,$ $K=0,$ $T=1,$ $%
R(0)=1.099,$ $R^{\prime}(0)=0,$ $\varphi(0)=0$

(a) thin line $\varphi^{\prime}(0)=0.3$

(b) heavy line $\varphi^{\prime}(0)=0.7.$

\includegraphics[
height=1.8in, width=1.8in, angle=270, trim = 0.1in 0.1in 0.1in 0.1in, clip
]{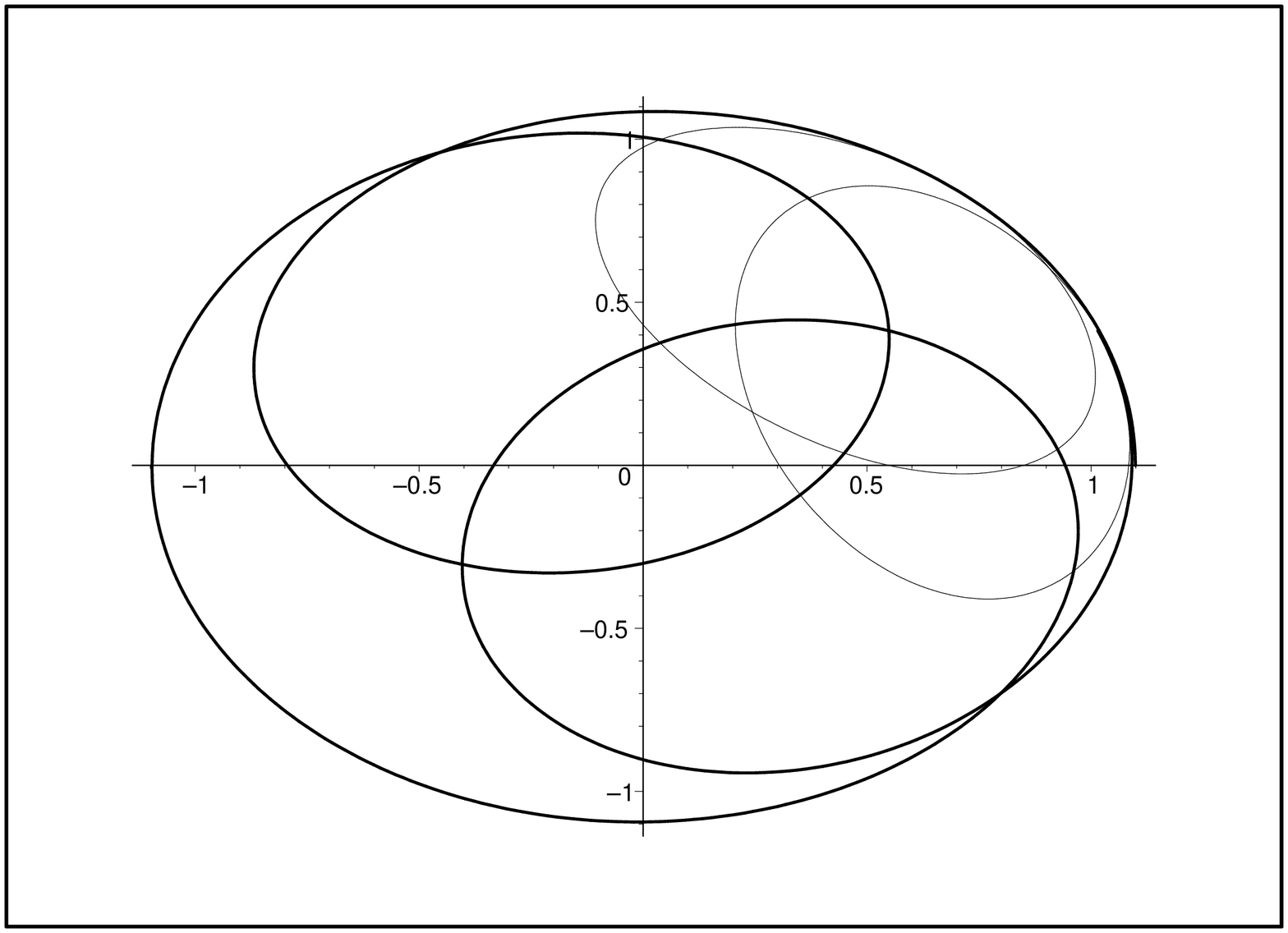}

FIG. 9. Amplitude-phase polar diagram for $n=4,$ $K=0,$ $T=1,$ $R(0)=1.099,$
$R^{\prime}(0)=0,$ $\varphi(0)=0$

(a) thin line $\varphi^{\prime}(0)=0.3$

(b) heavy line $\varphi^{\prime}(0)=0.7.$
\end{center}

\newpage

\section{Numerical solution for K}

Results of the numerical solution of our equations in the case of $K\neq0$
are presented in the figures 10 -- 15. The following important distinctions
from the case of $K=0$ must be mentioned:

(i) The periodicity of the spatial dependence is broken.

(ii) The frequency and amplitude modulation can be seen in Figs. 10 and 11.

(iii) A direction of the phase function staircase spatial dependence may
change to the opposite after some number of steps. The number of steps
between neighboring direction changes is random (see FIG. 15).

(iv) There exists a parameter range, where the trajectories in the
polar diagram show closed periodic movement: a synchronization
with the periodic contribution of the sine of the monotonously
increasing component of the phase function takes place.

\begin{center}
\includegraphics[
height=3in, width=2in, angle=270, trim = 0.1in 0.1in 0.1in 0.1in, clip
]{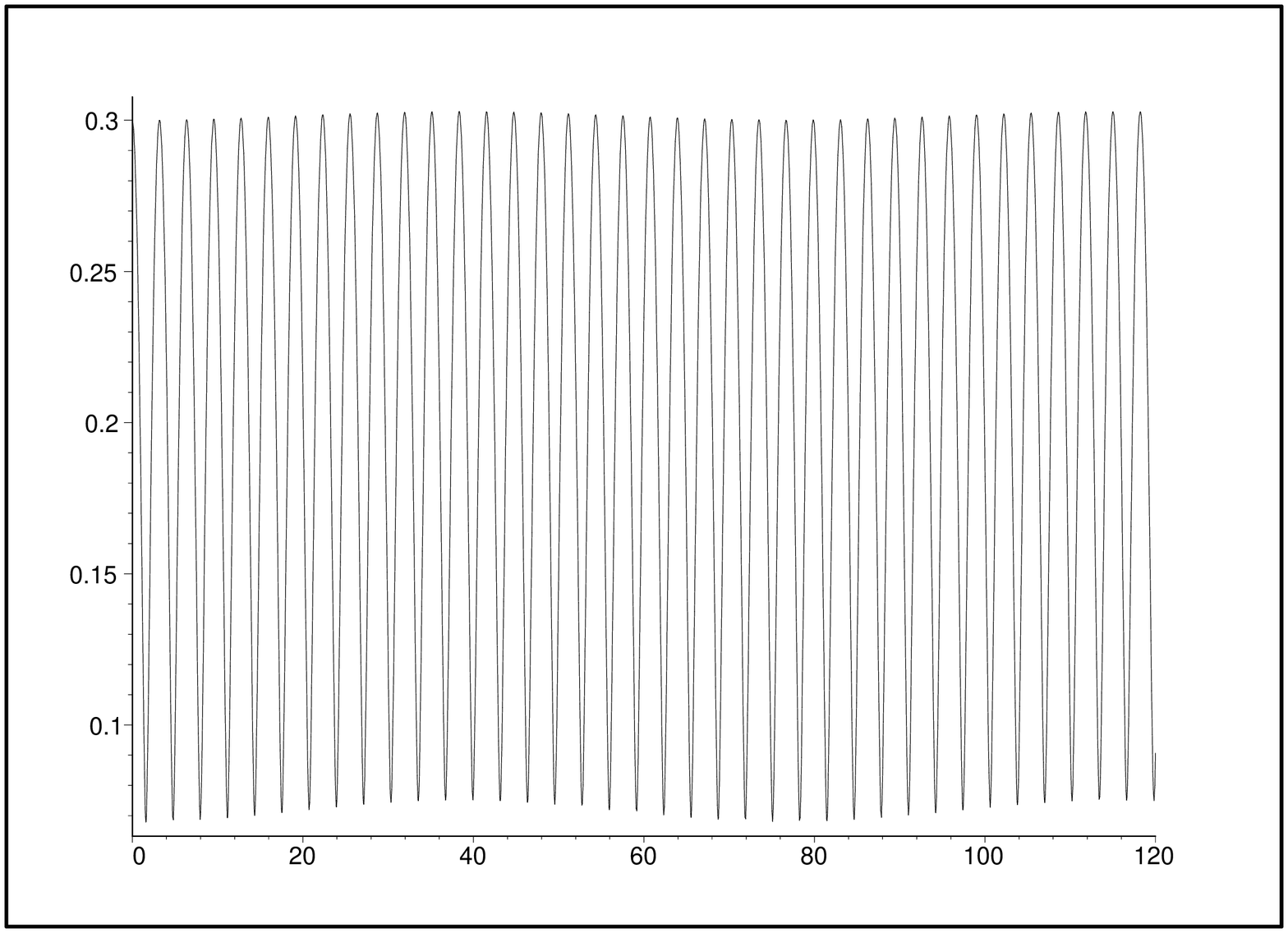}

FIG. 10. Spatial dependence of the amplitude function for $n=4,$ $K=1.6,$ $%
T=1,$ $R(0)=0.3,$ $R^{\prime}(0)=0,$ $\varphi(0)=0,$ $\varphi^{\prime
}(0)=0.3$

\includegraphics[
height=3in, width=2in, angle=270, trim = 0.1in 0.1in 0.1in 0.1in, clip
]{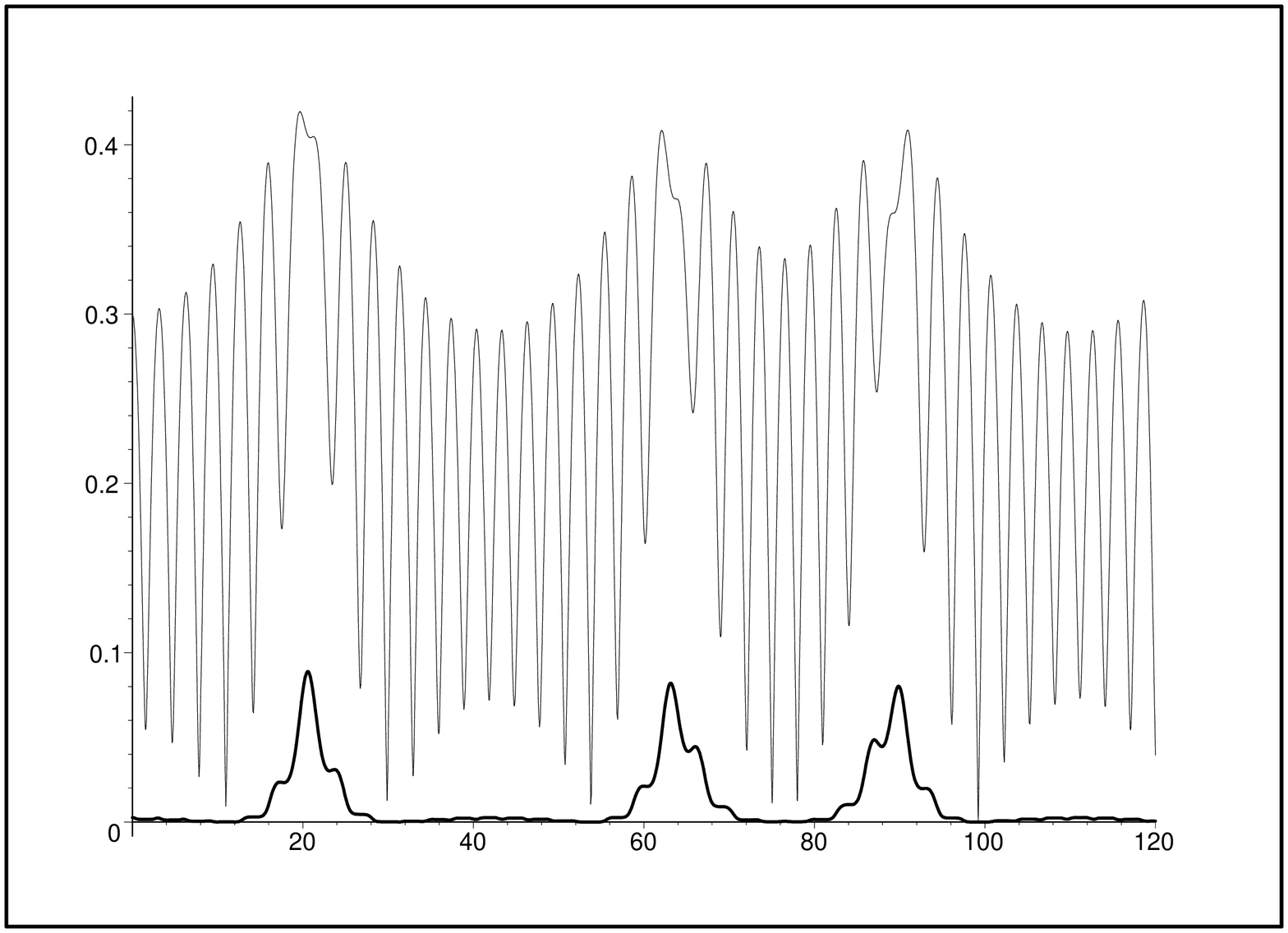}

FIG. 11. Spatial dependence of the amplitude function for $n=4,$ $K=1.6,$ $%
T=1,$ $R(0)=0.3,$ $R^{\prime}(0)=0,$ $\varphi(0)=0,$ $\varphi^{\prime
}(0)=0.75$: thin line

Spatial dependence of the parameter $C^{2}$ (see below) in arbitrary units:
heavy line.

\includegraphics[
height=3in, width=2in, angle=270, trim = 0.1in 0.1in 0.1in 0.1in, clip
]{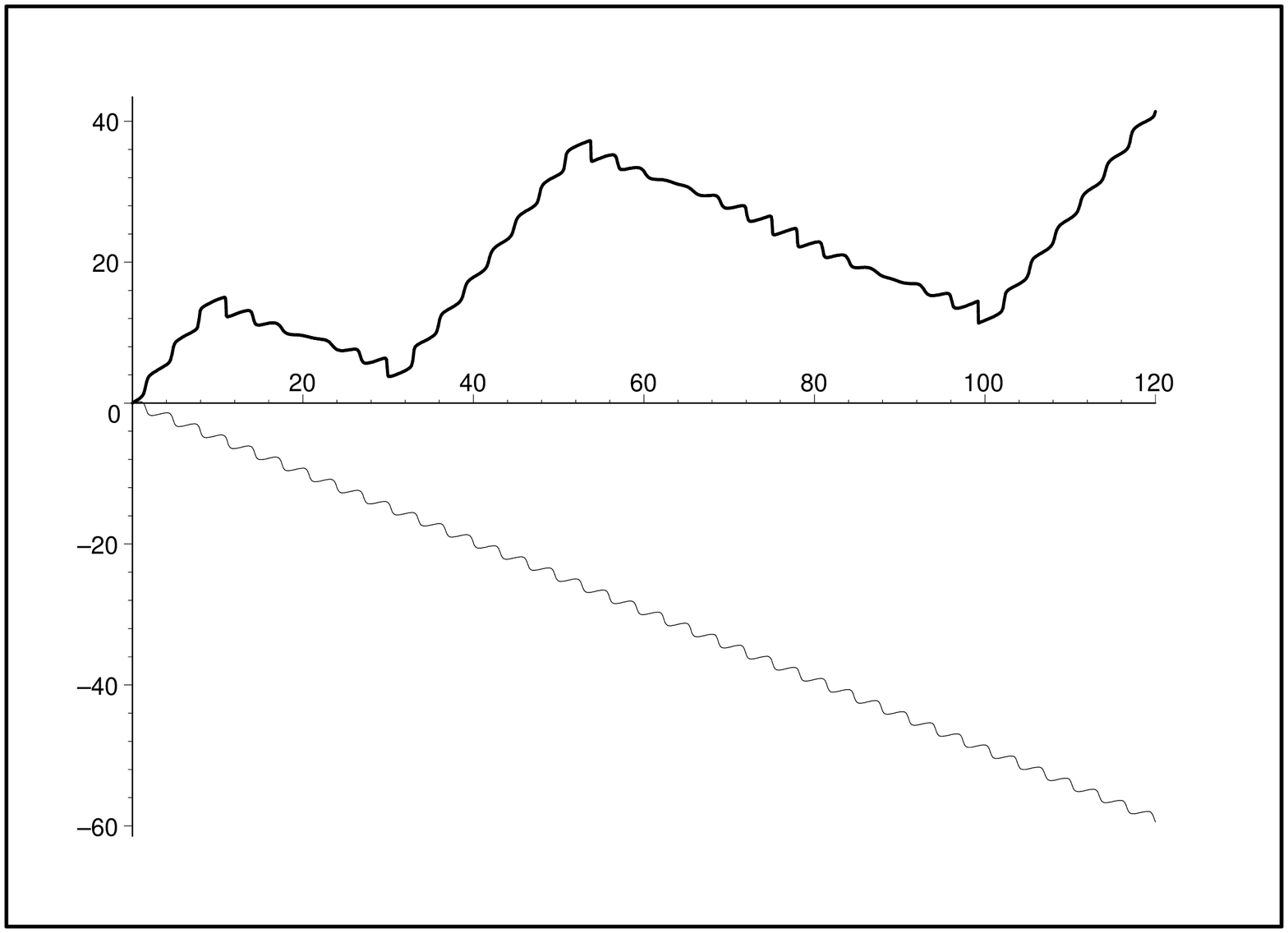}

FIG. 12. Spatial dependence of the phase function for $n=4,$ $K=1.6,$ $T=1,$
$R(0)=0.3,$ $R^{\prime}(0)=0,$ $\varphi(0)=0$

(a) thin line $\varphi^{\prime}(0)=0.3$

(b) heavy line $\varphi^{\prime}(0)=0.75.$

\includegraphics[
height=2.5in, width=2.5in,angle=270, trim = 0.1in 0.1in 0.1in 0.1in, clip
]{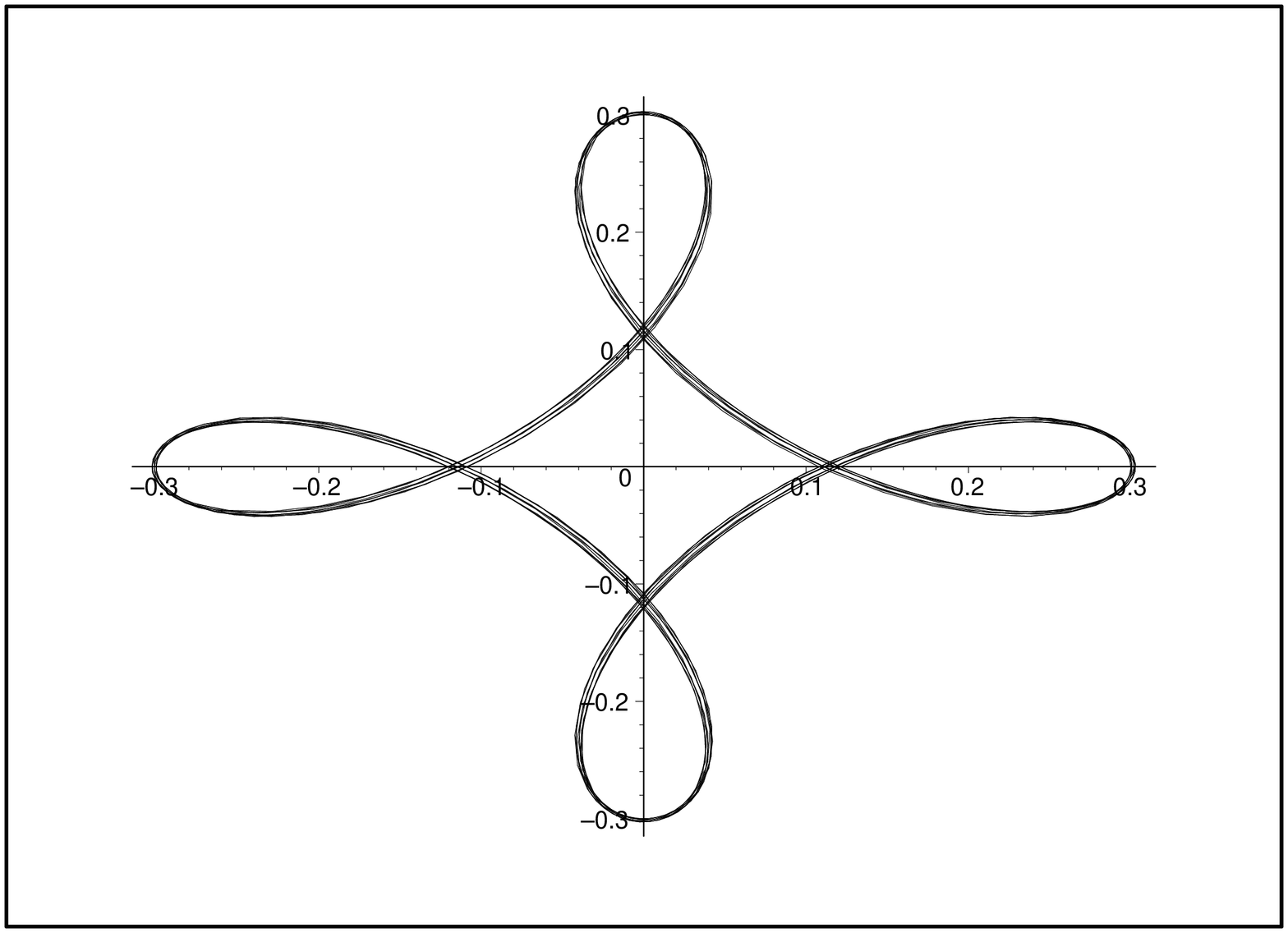}

FIG. 13. Amplitude-phase polar diagram for $n=4,$ $K=1.6,$ $T=1,$ $R(0)=0.3,$
$R^{\prime}(0)=0,$ $\varphi(0)=0,\varphi^{\prime}(0)=0.3$

\includegraphics[
height=2.5in, width=2.5in, angle=270, trim = 0.1in 0.1in 0.1in 0.1in, clip
]{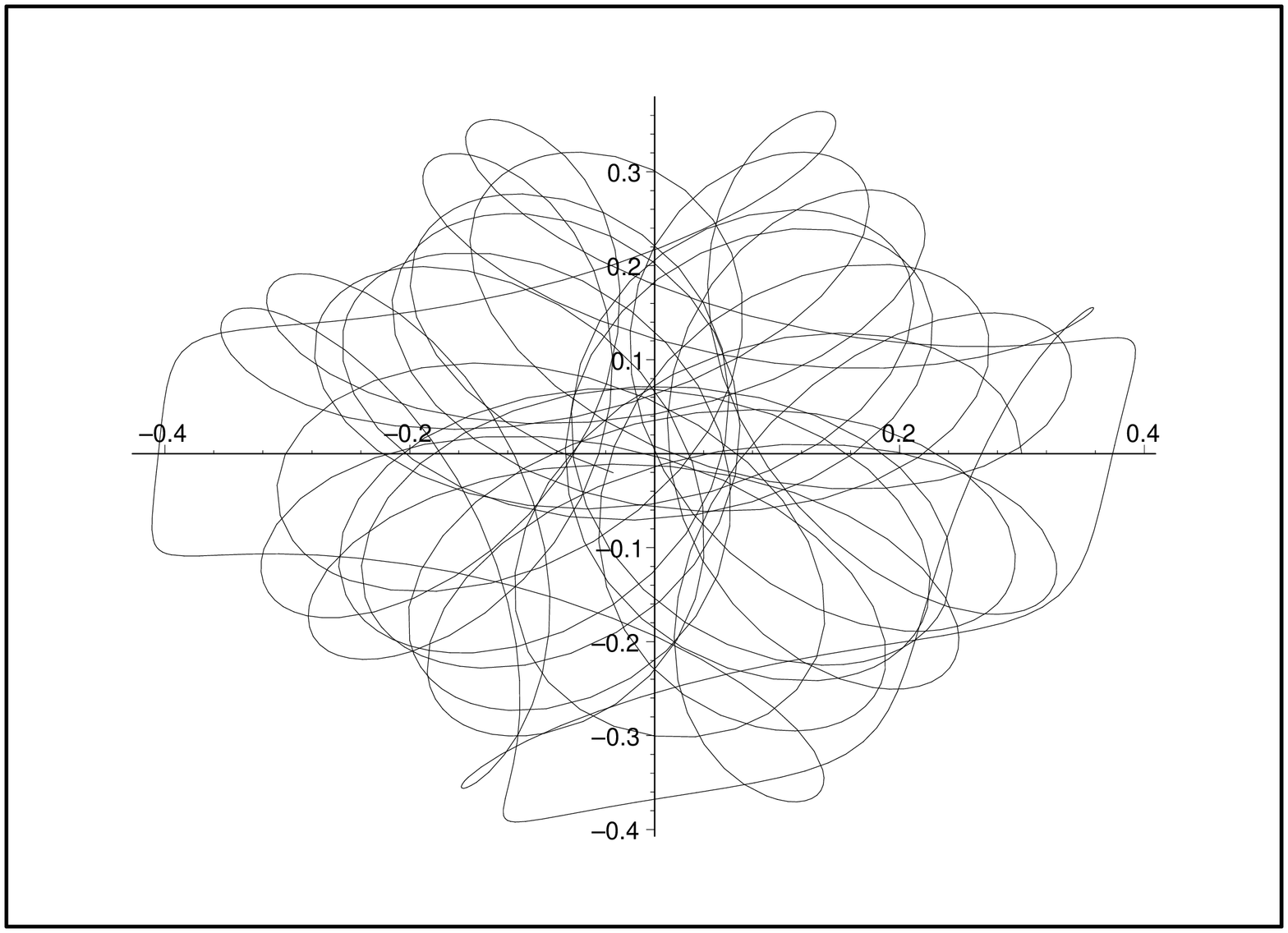}

FIG. 14. Amplitude-phase polar diagram for $n=4,$ $K=1.6,$ $T=1,$ $R(0)=0.3,$
$R^{\prime}(0)=0,$ $\varphi(0)=0,\varphi^{\prime}(0)=0.75$
\end{center}

\newpage

\section{Discussion}

Thus, the main distinctions of the solution in the case of the
non-vanishing anisotropy parameter $K$ are a random change of the
jumps direction (see FIG. 12) and an amplitude modulation of the
amplitude function (see FIG. 11). Just these random direction
changes lead to the tangled trajectory in polar coordinates
presented in FIG 14.

In the case of $K=0$ we introduced the integration constant $C_{0}$ (\ref%
{phasesolution}). Let us introduce the function:

\begin{equation*}
C(\xi)=\left[ \psi(\xi)-\frac{T}{2}\right] R(\xi)^{2}.
\end{equation*}

Direct differentiation shows that this function reduces to a constant in the
case of vanishing anisotropy parameter $K$:%
\begin{eqnarray}
\frac{dC}{d\xi}=2R\frac{dR}{d\xi}\left[ \psi-\frac{T}{2}\right] +R^{2}\frac{%
d\psi}{d\xi}=  \notag \\
=R^{2}(2\frac{R^{\prime}}{R}\varphi^{\prime} -T\frac{R^{\prime}}{R}%
+\varphi^{\prime\prime})=  \notag \\
=R^{2}(-R^{n-2}K\sin n\varphi)=-R^{n}K\sin n\varphi.
\end{eqnarray}

\begin{center}
\includegraphics[
height=3in, width=2in, angle=270, trim = 0.1in 0.1in 0.1in 0.1in, clip
]{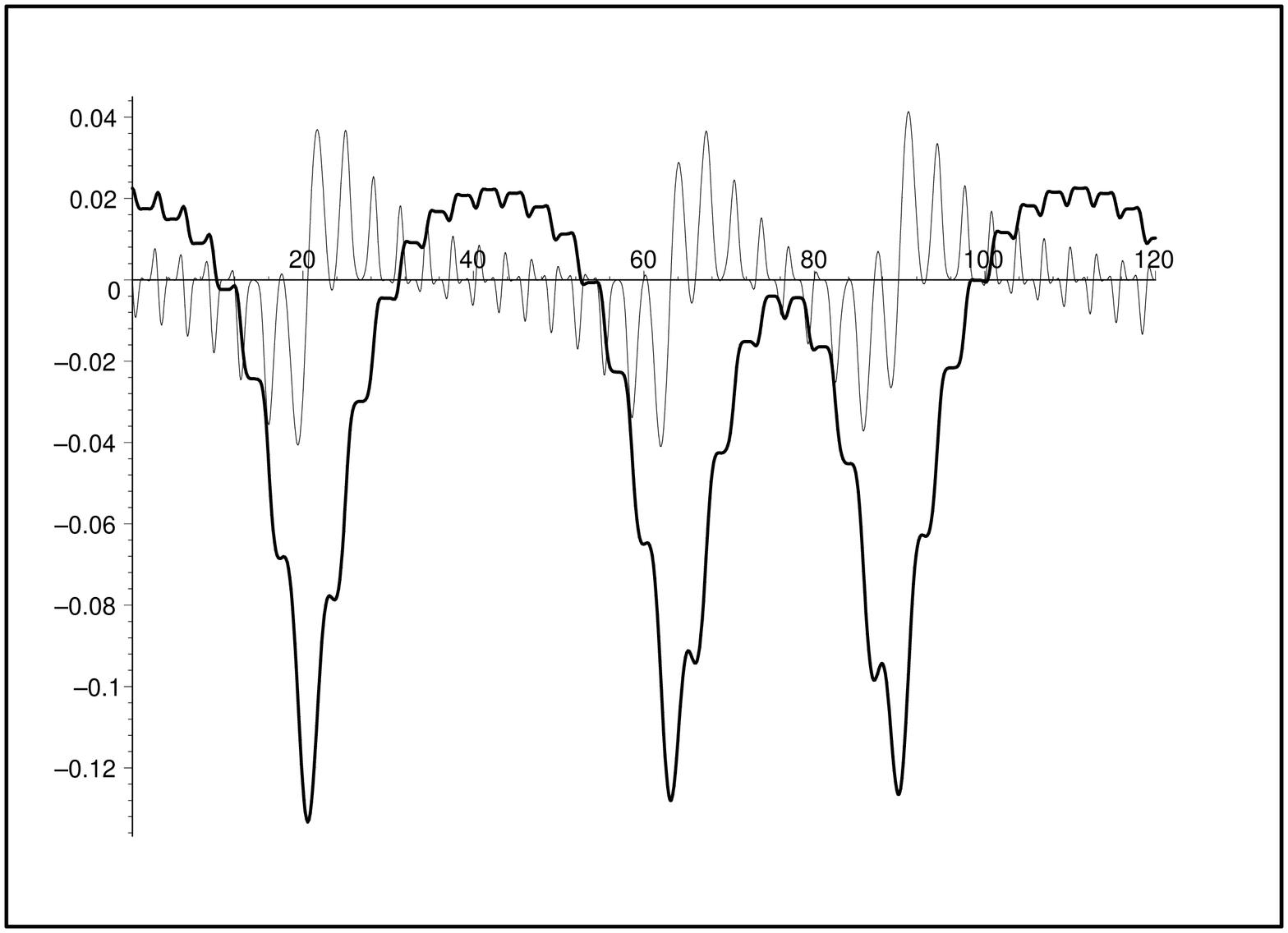}

FIG. 15. Spatial dependence of the parameter $C$ for $n=4,$ $K=1.6,$ $T=1,$ $%
R(0)=0.3,$ $R^{\prime}(0)=0,$ $\varphi(0)=0,\varphi^{\prime}(0)=0.75$

(a) thin line $C^{\prime}$

(b) heavy line $C.$
\end{center}

Comparison of FIG. 15 and FIG. 12 confirms that changes of the direction
take place in the points where $C(\xi)$ function changes its sign.

\section{Conclusion}

We have shown in this paper that the constant amplitude
approximation gives a poor description of the real picture of the
spatial evolution of the amplitude and phase functions for the
model of the incommensurate ferroelectric with the Lifshitz
invariant.


\begin {thebibliography}{99}

\bibitem {dzialoshin}{ I. Dzyaloshinskii, Soviet Physics JETP 19, 960
(1964).}

\bibitem{levaniuk}{A.P. Levaniuk and D.G. Sannikov, Fiz. Tverd. Tela 18,
423 (1976).}

\bibitem{hachatur}{ A.G. Khachaturian, \textit{Theory of Structural
Transformations in Solids, Wiley, 1983}}

\bibitem{char}{  E.V. Charnaya, S.A. Ktitorov, O.S. Pogorelova,
Ferroelectrics, 297, 29 (2003).}

\bibitem{izium}{ Ju. A. Iziumov, V.N. Syromiatnikov, Phase Transitions and
Crystal Symmetry, Moscow, Nauka, 1984.}

\bibitem{soliton}{Mark J. Ablowitz and Harvey Segur, Solitons
and the Inverse Scattering Transform, SIAM, Philadelphia, 1981}
\end{thebibliography}
\end{document}